\documentclass{nature_new}
\usepackage[normalem]{ulem}
\usepackage{graphicx}
\usepackage{dcolumn}
\usepackage{bm}
\usepackage{verbatim}
\usepackage[version=3]{mhchem}
\usepackage{amsmath}
\usepackage{hyperref}
\usepackage{caption}
\usepackage{color}
\usepackage{afterpage}
\usepackage[margin=10pt,font=small,labelfont=bf,labelsep=endash]{caption}

\title{Disordered Zero-Index Metamaterials Based On Metal Induced Crystallization}

\author{Henning Galinski$^{1}$, Andreas Wyss$^{1}$, Mattia Seregni$^{1}$, Huan Ma$^{1}$, Volker Schnabel$^{1}$, Alla Sologubenko$^{1,2}$, \& Ralph Spolenak$^{1}$}

\begin{document}

\maketitle

\begin{affiliations}
 \item Laboratory for Nanometallurgy, ETH Zurich, Zurich, Switzerland
 \item Scientific Center for Optical and Electron Microscopy, ETH Zurich, Zurich, Switzerland
\end{affiliations}

\begin{abstract}


Zero-index (ZI) materials are synthetic optical materials with vanishing effective permittivity and/or permeability at a given design frequency. Recently, it has been shown that the permeability of a zero-index host material can be deterministically tuned by adding photonic dopants. Here, we apply metal-induced crystallization (MIC) in quasi-random metal-semiconductor composites to fabricate large-area zero-index materials. Using Ag-Si as a model systems, we demonstrate that the localized crystallization of the semiconductor at the metal/semiconductor interface can be used as design parameter to control light interaction in such a disordered system. The induced crystallization generates new zero-index states corresponding to a hybridized plasmonic mode emerging from selective coupling of light to the \r{a}ngstr\"om-sized crystalline shell of the semiconductor. Photonic doping can be used to enhance the transmission in these disordered metamaterials as is shown by simulation. Our results break ground for novel large-area zero-index materials for wafer scale applications and beyond.

\end{abstract}

\section*{Introduction}
With the advent of optical metamaterials, numerous new functional materials emerged able to manipulate electromagnetic waves in an unprecedented manner. Controlled engineering of the electromagnetic properties, represented by the materials effective permittivity $\epsilon$ and permeability $\mu$, allowed for the experimental realization of materials with negative refractive indexes~\cite{Shelby77,PhysRevLett.95.137404}, flat optics based on plasmonic and all dielectric metasurfaces~\cite{Yu333,Chen2018} and even materials with their real permittivity close to zero~\cite{Maas2013,Liu2008,Adams2011,Caspani2016,Kim:16}. Intriguingly, a vanishing permittivity or permeability in such zero-index materials leads to new physical regimes of light interaction with matter~\cite{Engheta2013}. Zero-index (ZI) materials exhibit quasi-static behaviour at optical frequencies, as both the phase velocity $v_p=c/\sqrt{\epsilon'}$ and the wavelength $\lambda$ tend to infinity as the real part of the permittivity $\epsilon'$ approaches zero.\\   
\\
Despite the "quasi-static" nature of the fields within the ZI material, the dynamics of the field are still that of a transverse wave~\cite{Ziolkowski2004}. The homogenized electromagnetic fields and the high impedance mismatch $Z=\sqrt{\mu/\epsilon}$ in ZI materials led to a considerable theoretical and experimental interest in this matter. Many unconventional phenomena have been predicted by theory including supercoupling of light through arbitrarily shaped waveguides~\cite{Silveirinha2007}, tunneling of light through subwavelength channels~\cite{Silveirinha2007,Silveirinha2006,5951087}, wavefront shaping~\cite{Alu2007,7342497}, electric levitation of nanoscale emitters~\cite{PhysRevLett.112.033902}, enhancement of nonlinear optical effects~\cite{Argyropoulos2012}, polarization conversion~\cite{Ginzburg2013} and coherent prefect absorption~\cite{PhysRevB.86.165103}.\\
\\
Quite interestingly, even natural materials, such as metals exhibit ZI behaviour, typically situated in the UV. This behavior is easily understood by considering the Drude Model~\cite{Mizutani}, which describes the motion of the conduction electrons within a metal stimulated by an external time-harmonic field. On resonance, the real part of the dielectric response $\epsilon'\approx\epsilon_{c}-\omega_{p}/\omega^2$ vanishes when the oscillating term cancels the contribution of the positively charged metal ion cores, i.e. $\omega=\omega_{p}/\sqrt{\epsilon_{c}}$. In the vicinity of this resonance ZI behavior can be expected.\\
\\
A particular interesting route in ZI material research, is the idea to embed nonmagnetic inclusions in an epsilon near-zero (ENZ) host medium. This concept developed in a seminal work by Silveirinha, Liberal and Engheta~\cite{Silveirinha2007a,Liberal1058} enabled engineering of the permeability $\mu$ of an ENZ material by only changing the geometric shape or dielectric constant of the introduced inclusions. 
In this scenario, materials with both $\epsilon$ and $\mu$ near zero (EMNZ) can be engineered and the impedance mismatch between the material and electromagnetic (EM) radiation from free space is eliminated. Fully near-zero materials are vacuum fluctuation free~\cite{Liberal822,PhysRevA.96.022308} and can serve as a new platform for quantum optics and quantum information. However, a major challenge is the problem of limited scalability. Typically, metamaterials are fabricated using lithography techniques and cannot be scaled above the gram level. It is therefore highly desirable to find new approaches that can transform the conceptional breakthroughs in zero-index media to wafer scale applications.\\
Here, we use disordered metasurfaces as a new design approach for large-scale zero-index media. While usually lacking absolute geometric control, disordered random media can be easily realized on large scales, are polarization independent and are known to harbour exciting phenomena~\cite{Wiersma2013}, ranging from wavefront shaping~\cite{Jang2018}, lasing~\cite{Wiersma2008}, photocatalysis~\cite{Tian2017}, perfect absorption~\cite{ADOM:ADOM201600580}, light trapping~\cite{Lee8734}, circular dichroism~\cite{doi:10.1021/acsphotonics.7b01460} to tunable structural coloration~\cite{LSA2017}. A simple mechanism to engineer a disordered material is phase separation of immiscible phases. When two immiscible materials are mixed together, e.g. oil and water, they instantaneously tend to form separate phases. In solids, this process of phase separation leads to a spatial reorganization of the system, typically accompanied by the formation of quasi-random nanometric patterns.\\
Considering the special case of an immiscible pair of metal and semiconductor (see examples in Fig.~\ref{fig:1}(a)), the difference in crystallization temperature of the two materials can lead to a scenario, where the metal phase after phase separation is crystalline while the semiconductor phase stays amorphous. To this effect, it has been recently shown that similar to sintering aids in ceramics, the presence of a metal in an amorphous semiconductor reduces the crystallization temperature $T_c$ and enables preferential crystallization of the semiconductor at the metal/semiconductor interface.\\
This phenomenon is illustrated in Fig.~\ref{fig:1}(b) and commonly dubbed ''metal induced crystallization'' (MIC)~\cite{bookMIC,doi:10.1002/adem.200800340}. As we show below, such spatially localized crystallization furnishes a precise way to engineer the thickness of the crystalline layer by thermal treatments resulting in a quasi-random structure with core-shell features reminiscent of nanomatryoshkas~\cite{doi:10.1021/jp9095387} or concentric nanorings~\cite{doi:10.1021/acsphotonics.5b00133}.\\

\section*{Zero-Index Phase Diagrams}
\begin{figure}[t!]
    \centering
    \includegraphics[width=\textwidth]{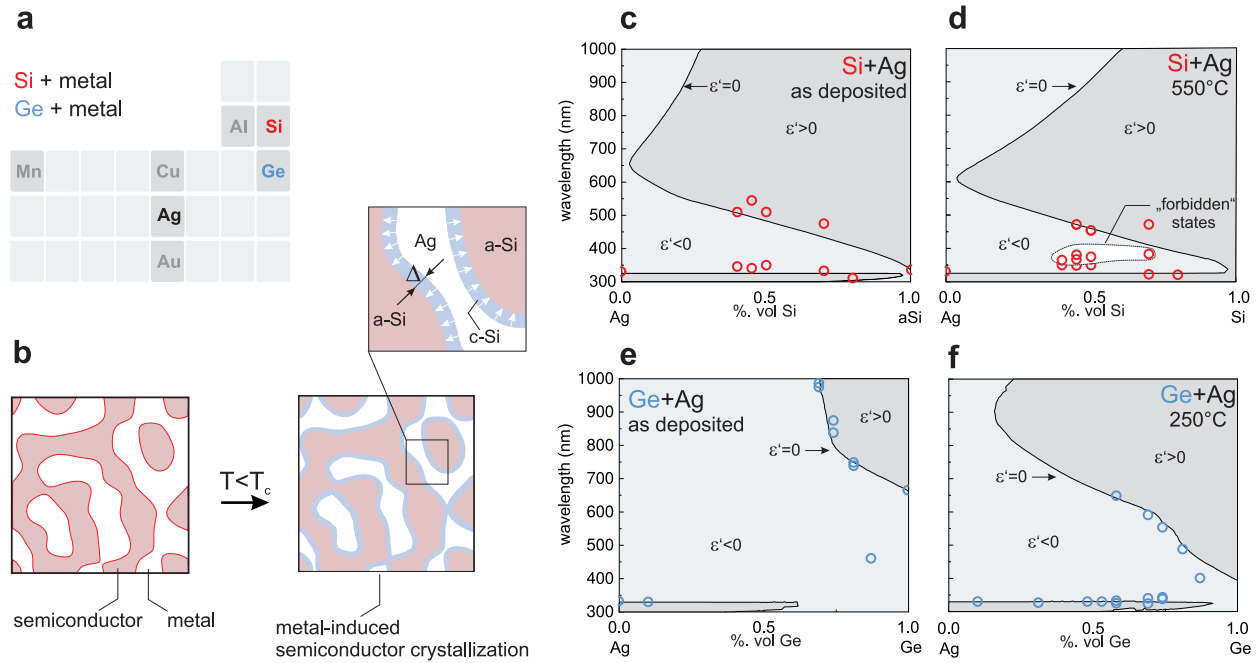}
    \caption{Designing zero-index materials by metal induced crystallization. (a)  Selection of immiscible metal/semiconductor pairs (b) Schematic of metal induced crystallization in an immiscible metal/semiconductor metamaterial starting from a "phase separated" state with an amorphous (red) semiconducting phase and crystalline metallic phase (white). After thermal treatment the semiconductor crystallizes (blue) at the metal/semiconductor interface forming a crystallized layer with thickness $\Delta$. Panels (c)-(f) report near-zero phase-diagrams of the as-deposited and an arbitrary chosen annealed state for Ag-Si (c)-(d) and Ag-Ge (e)-(f), respectively. Experimentally determined $\epsilon'=0$ states are marked by red (Ag-Si) and blue (Ag-Ge) circles, while the two phases of $\epsilon'>0$ and $\epsilon'<0$ given by an effective medium approximation are coloured in light-blue and grey. It is to note that in (d) ``forbidden`` states emerge, which go beyond the employed effective medium description.
    }
    \label{fig:1}
\end{figure}
In our experiments, we start with Ag-Si and Ag-Ge as specific examples of an immiscible metal-semiconductor pair (Fig.\ref{fig:1}a). Ag-Si and Ag-Ge thin films of $50$~nm thickness were co-deposited via magnetron co-sputtering and successive annealing was performed in order to induce metal induced crystallization (MIC). Based on ellipsometric measurements, we have derived phase diagrams indicating phases with metallic ($\epsilon'<0$) and dielectric ($\epsilon'>0$) behaviour. The phase-diagrams are based on experimentally determined optical constants of the single elements (see Supplementary Information). Thereby, a simple effective medium approximation (EMA) (see Supplementary for detailed information) is used to estimate the complex permittivity of the metamaterials and derive the phase diagram. Figure~\ref{fig:1} (c)-(f) report different $\epsilon'$-composition phase-diagrams for the as-deposited state and one selected annealing state. In each phase diagram, the experimentally measured ENZ-states of the metamaterial are marked for comparison. The phase diagrams are in good agreement with the experimental measurements and reveal ENZ states in the UV and red. Quite remarkably, we also observe the formation of ``forbidden'' ENZ states, i.e. states not captured by the EMA, in case of annealed Ag-Si metamaterials (see Fig.~\ref{fig:1}(d)). Such states are formed for various compositions between $365$ and $384$~nm. As a good zero-index material should satisfy $\epsilon'\approx0$ and as well exhibit low optical losses, i.e. $\epsilon''<1$, we will concentrate on the Ag-Si systems. Detailed information on the Ag-Ge system can be found in the supplementary information.

\section*{Light-Matter Interaction}
\begin{figure}[t!]
    \centering
    \includegraphics[width=1.07\textwidth]{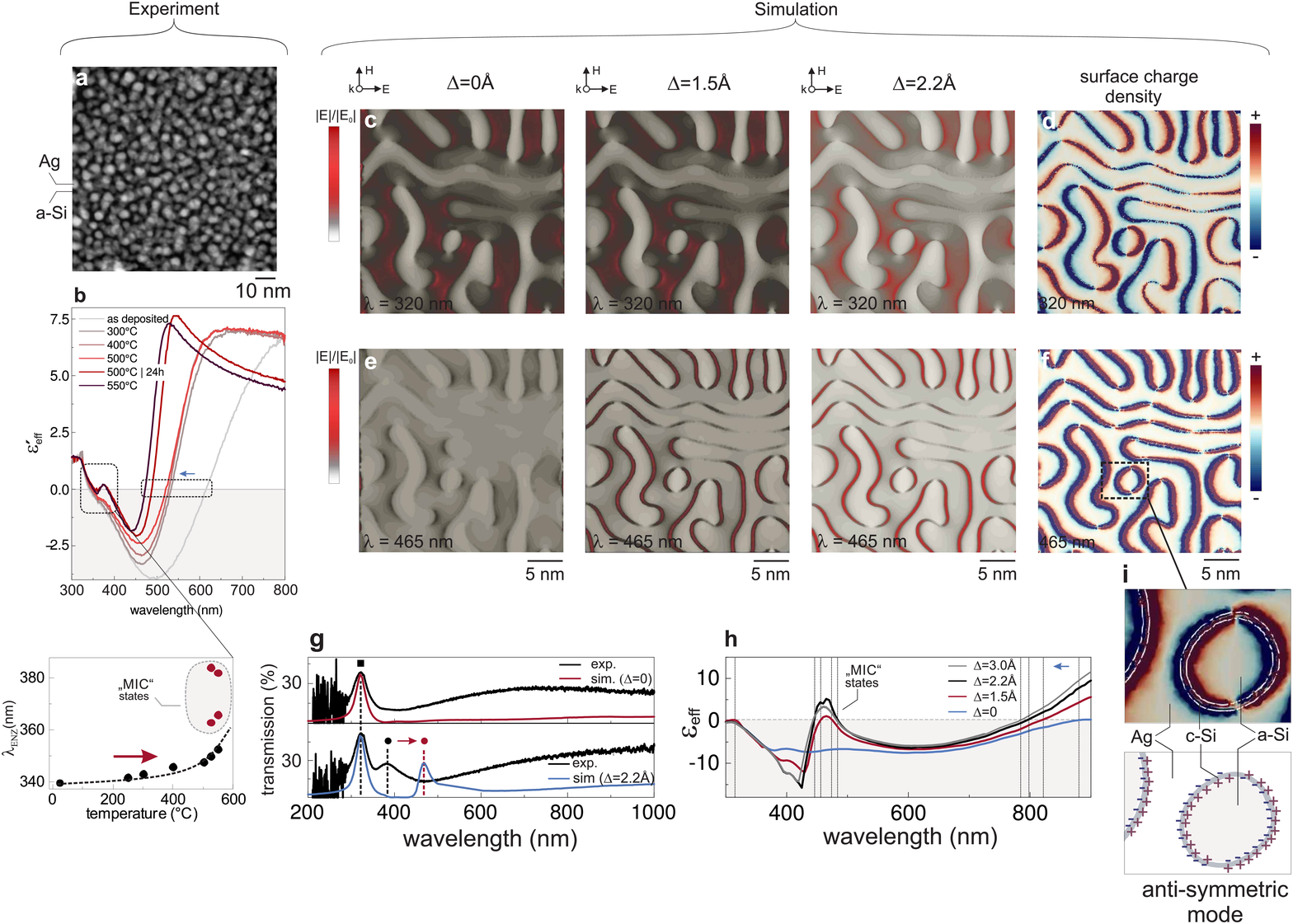}
    \caption{Optical properties and light interaction. Panel (a) shows a STEM image of an as-deposited Ag$_{.55}$Si$_{.45}$ thin film  interpenetrating Si (black) and crystalline Ag (grey) phases. (b) Real part $\epsilon'_{\text{eff}}$ of the dielectric function of Ag$_{.55}$Si$_{.45}$ as function of thermal treatment. The annealing states ($500$ and $550^\circ$C) exhibit "MIC" based ENZ states. (c)-(e) Simulated electric field enhancement for a $50$ nm thick Ag$_{.55}$Si$_{.45}$ thin film illuminated by a plane wave with the electric field polarized along the x axis at $\lambda{_0}=320$~nm (c) and $\lambda{_0}=465$~nm (e). Each panel shows the systems response for a crystalline Si phase of different thickness $\Delta$. In case of (c) maximal field enhancement ($4\times$) is achieved between adjacent Si-domains, while in (e) maximal field enhancement ($7\times$) is achieved by near-field localization of light in the $2.2${\AA} thick crystalline silicon shells. (d)-(f) Surface charge density of Ag$_{.55}$Si$_{.45}$ metamaterial with $2.2${\AA} crystalline shell thickness. (d) The surface charge density distribution is indicative of a collective dipolar mode. (e)-(i) High charge concentration at the Ag/c-Si and c-Si/a-Si interfaces, causing the formation of an hybridized anti-symmetric mode. (g) Comparison between experimentally measured and simulated transmission spectrum for $20$~nm thick metamaterials (h) Retrieved real part of the permittivity of simulated Ag$_{.55}$Si$_{.45}$ metasurfaces with different c-Si shell thickness $\Delta$.}
    \label{fig:2}
\end{figure}

We study the physical mechanisms responsible for the occurrence of these ´´forbidden´´ ENZ states, by analyzing one selected composition, namely Ag$_{.55}$Si$_{.45}$, of the Ag-Si metamaterial in detail. Figure~\ref{fig:2}a shows a scanning transmission electron microscopy (STEM) micrograph depicting the interpenetrating amorphous Si (black) and crystalline Ag (grey) phases of the metamaterial, confirming the "phase-separated" state of the nano-composite after deposition. The optical properties, in terms of permittivity, have been determined using ellipsometry. Figure~\ref{fig:2}(b) presents the effect of  annealing on the real permittivity of this nano-materials. The spectral response can be subdivided in two spectral regions with characteristic ENZ behavior.
\afterpage{\clearpage}
 The first ENZ states, located in the UV and violet combine ``forbidden'' states and the known Drude-Lorentz response of silver~\cite{PhysRevB.91.235137}. Latter feature is only weakly dependent on compositional changes and characterized by a red-shift of the critical ENZ wavelength with increasing annealing temperature (Fig.\ref{fig:2}b). This trend is easily explained by an on-average increasing dielectric constant of the semiconducting phase. At temperatures $\geq 500^\circ$C, a new ´´forbidden´´ state emerges. This state is particularly interesting, as it is controllable by thermal treatment offering a new way to deterministically control the optical response in an overall random and disordered system.\\
\\ 
In the second region, located in the red and near IR, the spectra show a rich dynamical scenario as the annealing temperature is increased. The critical ENZ-wavelength is continuously blue-shifted to lower wavelength. However, it has to be noted that the losses in this region (see Supplementary Information), represented by the imaginary part of the permittivity $\epsilon''$, within the designed materials are still too high to allow for ENZ behavior. This effect is mostly due to the presence of low quality amorphous Si, characterized with high intrinsic losses with respect to its crystalline phase. However, these quality issues can be overcome by resorting to ion-beam sputtering or ion-beam-assisted sputtering~\cite{PhysRevD.92.062001}.\\
\\ 
In order to interpret the experimental results, we realized a series of three-dimensional Finite-Element Method (FEM) full-wave simulations. We want to stress that our numerical simulations are not intended to replicate the experimental results one-to-one but to derive an intuitive physical picture of the light-matter interaction. To this extent, we built a simplified model of a phase-separated composite using the phase-field method~\cite{ZHU2016322}. Our model reproduces the volume fraction and compositional domains of our samples as determined by Scanning Transmission Electron Microscopy, see Figure~\ref{fig:2}(a). The metal-induced crystallization is modelled by assuming a inward-growing conformal layer on the Si-phase. More details on the FEM simulations can be found in the supplementary online text.\\
Figure~\ref{fig:2}(c) and (e) report the local field enhancement at two ENZ wavelength, namely $320$~nm and $465$~nm for different crystalline Si (c-Si) shells. We observe, that the coupling of light to the ZI material at these two wavelength is entirely different. At $320$~nm, corresponding to the Drude response, the local electric field enhancement is confined to the nanometer-scale Ag-phase due to capacitive coupling between different regions of the Si-phase independent of the c-Si-shell thickness. The coupling abruptly changes at $465$~nm, where light is tightly confined to the \r{a}ngstr\"om-sized c-Si shell of the metasurface. We have applied effective parameter retrieval~\cite{PhysRevE.71.036617} to assess the effective permittivity of the simulated ZI materials. The quantitative prediction of our FEM model shown in Fig.\ref{fig:2}(h) reproduces the formation of the ``MIC-based'' near-zero state and exhibits all characteristic features also found in the experiment, as shown in Fig.\ref{fig:2}(b). The observed red-shift between the simulated and experimental near-zero states can be attributed to the known influence of particles' aspect ratio on plasmonic resonances~\cite{maier2007plasmonics}. Due to the used periodic boundary condition in the simulation, we artificially truncate the phases of our material and thus systematically underestimate their aspect ratio.\\
\\
To further analyze the properties of ´´MIC-based´´ near-zero states, we mapped in Figure~\ref{fig:2}(d) and (f) the surface charge density in the case of $\Delta=2.2\AA$. While the near-zero state at $320$~nm is characterized by classical collective plasmonic dipolar mode of the material with quasi uniform field in the semiconductor, the ``MIC-based'' state emerges due to coupling of local anti-symmetric modes, where charges primarily reside on the inner and outer-interfaces of the crystalline Si-shell (Fig~\ref{fig:2}(h)). In the experiment as well as in the simulation, this mode causes enhanced transmission in the far field as shown in Fig.~\ref{fig:2}(g). This phenomenon can be understood in the context of optical cloaking in core-shell nanoparticles~\cite{Kerker:75,doi:10.1080/02786828208958594,PhysRevE.72.016623,Rohde:07,Muehlig2013}. As analyzed by Kerker~\cite{Kerker:75,doi:10.1080/02786828208958594} and others~\cite{PhysRevE.72.016623,Rohde:07}, the polarizibility, i.e the scattering of concentric spheres of properly chosen permittivity ($\epsilon_\text{core}<\epsilon_\text{medium}<\epsilon_\text{shell}$), becomes zero for a critical shell thickness, when the scattered waves produced by the dipolar resonances of the inner and outer shell destructively interfere. Invisibility, i.e. the complete suppression of scattering of light in this frequency range, is achieved for specific ratios of the core $r_\text{core}$ and total radius $r$ of the particle~\cite{doi:10.1080/02786828208958594,PhysRevE.72.016623},
\begin{equation}
    \frac{r_\text{core}}{r}
    =\sqrt[3]{
    \frac{(\epsilon_\text{shell}-\epsilon_\text{medium})(2\epsilon_\text{shell}+\epsilon_\text{core})}{(\epsilon_\text{shell}-\epsilon_\text{core})(2\epsilon_\text{shell}+\epsilon_\text{medium})}.
    }
\end{equation}
Here, we create a similar phenomenon in a disordered system with significant optical losses. Modulating the sub-nanometer thickness of the crystalline Si-shell by temperature, generates a hybridized ''ENZ´´-resonance state that effectively couples incoming radiation to the crystalline Si-shell. These local plasmonic resonances do not propagate into the far-field and therefore successfully suppress the materials scattering response. It is to note, that the experimental realization of such highly confined modes represents a new practical way to study the extreme coupling regime, where the classical model of electron response is known to fail~\cite{Ciraci1072}.

\begin{figure}[t!]
    \centering
    \includegraphics[width=\textwidth]{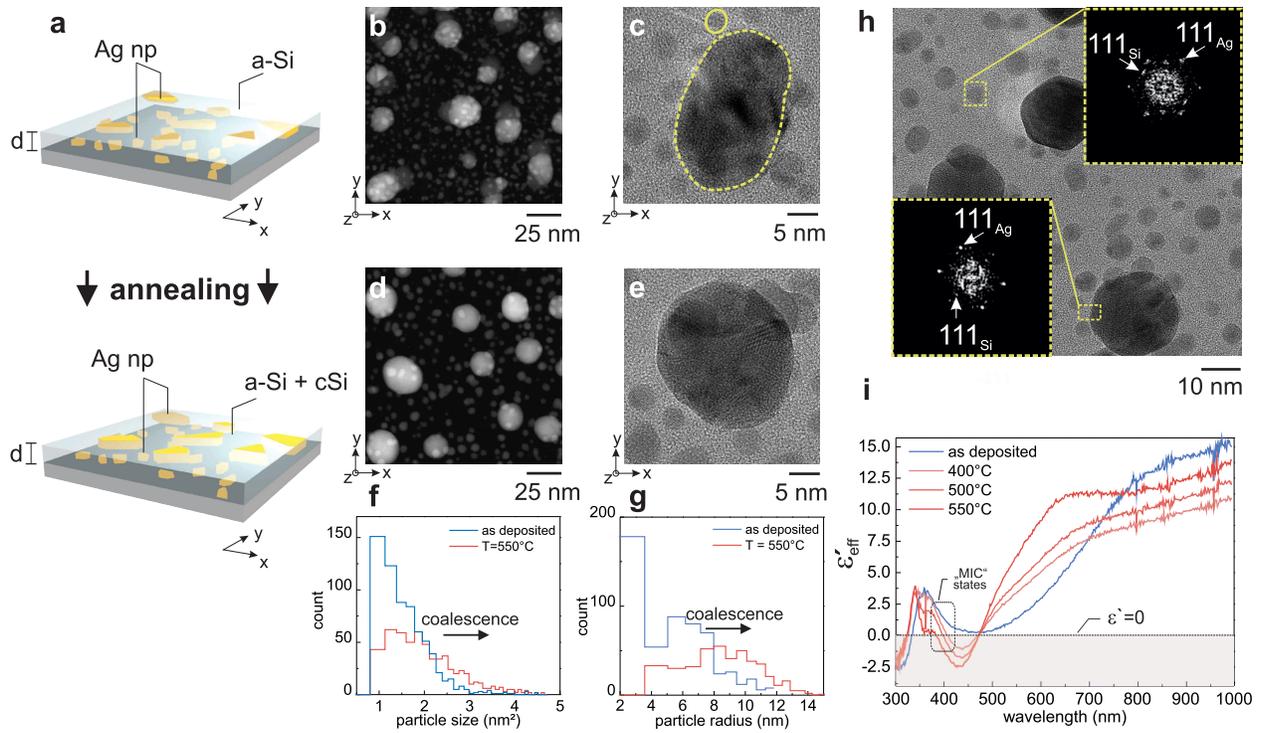}
    \caption{Optical properties and morphology of Ag-Si based ENZ materials by metal induced crystallization. (a) Schematic presenting the morphology of immiscible Ag-Si thin films before and after annealing as determined from TEM measurements (b)-(e) STEM and TEM micrographs of a pristine Ag$_{.30}$Si$_{.70}$ sputtered onto a SiN$_x$ membrane (b)-(c) and after in-situ annealing at $550^\circ$C for $20$~min (d)-(e). Panels (f)-(g) compare the particle size distribution of small (f) and large (g) Ag nanoparticles before and after annealing at  550$^\circ$C.
    (h) High-resolution TEM micrograph of the in-situ annealed specimen with a corresponding Fast Fourier Transforms acquired from the marked areas at Ag particles. (i) Real part of calculated dielectric function of Ag$_{.30}$Si$_{.70}$ as function of thermal treatment.}
    \label{fig:3}
\end{figure}

\section*{In-situ Metal Induced Crystallization}
To study and validate the crystallization of semiconducting phase domains adjacent to the metallic phase in the experiment, we resort to \textit{in-situ} transmission electron microscopy. Figure \ref{fig:3} (b)-(c) show a metal-dielectric composite composed of subwavelength Ag nanoparticles (Ag-nP) embedded in the amorphous Si-matrix. The Ag-nP population exhibits a bimodal size distribution stable against annealing up to $550^\circ$C as is confirmed by direct observation during in-situ heating in TEM. Only marginal coarsening and particle growth was detected by the post-mortem analysis of the micrographs (Fig.\ref{fig:3}(d)-(e)).\\
\\
Fast-Fourier Transform (FFT) analysis of the micrographs acquired during the in-situ heating experiments evidenced the occurrence of the crystalline Si-phase exclusively at the perimeters of small and large Ag-particles (see Figure \ref{fig:3}h). Such an occurrence was never detected in the non-heat treated material. The localized crystallization of Si at the Ag-particles correlates with the observed optical response, due to the formation of “MIC-based” states (Figure ~\ref{fig:3}(i)).

\section*{Photonic Doping}
\begin{figure}[t!]
    \centering
    \includegraphics[width=\textwidth]{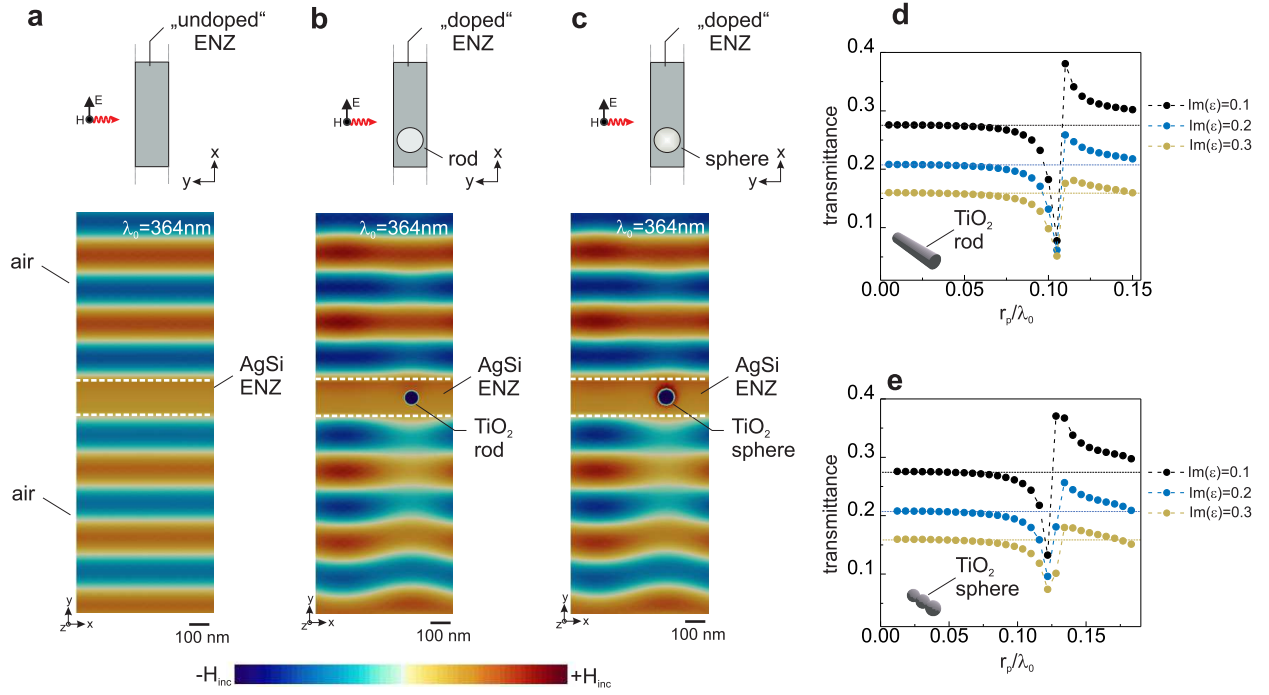}
    \caption{Photonic Doping (a) Schematic of the undoped Ag-Si ENZ-host and simulated magnetic field magnitude distribution for a $150$ nm thick ENZ slab 
    illuminated by a plane wave with the magnetic field polarized along the z axis at $\lambda{_0}=364$~nm. (b) and (c) same as in (a), but the ENZ slab has been doped with a dielectric TiO$_2$ rod (b) and dielectric TiO$_2$ particles ($\epsilon=13.4$) (c). Panel (d)-(e) show the magnitude of the transmission coefficient as function of the rod (d) or particle (e) radius $r_p/\lambda_0$ for three different loss values $Im(\epsilon)$. In both case, the overall transmission can be enhanced by using either a rod or particle once the radius $r_p$ reaches a critical size.}
    \label{fig:4}
\end{figure}

As demonstrated above, a dynamic optical response with $\epsilon\approx0$ can be created by exploiting the concepts of phase separation and MIC in metal-semiconductor mixtures. An intriguing question concerns the possibility of reducing material-specific losses in these systems to design a class of metamaterials with $\epsilon\approx0$ and  $\mu\approx0$ behavior. As shown recently by Liberal \textit{et al.}, $\epsilon$ and $\mu$ near zero behavior, can be realized by photonic doping, i.e. the introduction of dielectric nano-rods or similar in the material. In order to theoretically study the possibility of "photonic doping" in MIC-based ENZ materials, we realized a series of finite-element (FE) simulations. We employed two-dimensional simulations based on the measured permittivity of a Ag$_{.55}$Si$_{.45}$ thin film annealed at $550^\circ$C. Figure~\ref{fig:4}(a) reports the interaction of a TM plane wave under normal incidence with a $150$ nm Ag$_{.55}$Si$_{.45}$ thin film. 
The computed magnetic field shown in Fig.\ref{fig:4}(a) exhibits a constant magnetic field within the Ag$_{.55}$Si$_{.45}$ thin film confirming characteristic ENZ behaviour. Doping the ENZ host material with a dielectric rod and dielectric particle made of TiO$_2$, see Fig.~\ref{fig:4}(b)-(c), results in an either enhanced or highly reduced transmission as is shown Fig.~\ref{fig:4}(d)-(e). This is a very important result, as it confirms that the MIC-based thin films can serve as ENZ host material whose material properties can be further modulated by photonic doping. Quite interestingly, the optical response of the doped ENZ material exhibits unique features of a Fano resonance~\cite{Limonov2017}. In particular, the field distribution and the fact that the optical response can be switched by geometrical means from high to low transmission are reminiscent of a Fano response in dielectric resonators~\cite{PhysRevB.95.165119}.\\

\section*{Conclusion}

In this article, we have designed, fabricated and characterized disordered optical metamaterials with vanishing permittivity consisting of immiscible metal-semiconductor pairs. We have demonstrated that metal-induced crystallization can be used to engineer a set of new near-zero states in such disordered quasi-random metamaterials. These near-zero states emerge due to the formation of a temperature controllable \r{a}ngstr\"om-sized crystalline silicon (c-Si) shell in the Ag-Si metamaterial. We show that near zero-states are result of a collective resonance, where light is confined to the \r{a}ngstr\"om-sized c-Si shell by a mechanism analogous to optical cloaking in concentric core-shell nanoparticles. This demonstration of controlled coupling of light to \r{a}ngstr\"om-sized volumes in a disordered system represents a tremendous opportunity to study the extreme-coupling regime in robust and scalable disordered optical media. Using simulations we demonstrate, that the near-zero response of such a large-area metamaterial can be further modulated by photonic doping to create a full zero-index response. As, the proposed disordered metamaterials are fabricated by simple deposition methods and are not constrained with respect to their active area, we expect that these ideas can be exploited to design zero-index media for wafer scale-applications for integrated photonics, thermal sensors and quantum information.

\section*{Methods}
\subsubsection*{Fabrication}

Ag-Ge and Ag-Si ZI films of different composition were deposited by magnetron sputtering (PDV Products).
All films were deposited on silica glass under a controlled Argon atmosphere of $50 $~mTorr, with a $50$~sccm flow. The substrate was kept under a constant rotation of $30$~rpm to guarantee a homogeneous composition over the whole substrate. The sputter sources were set to $35^\circ$ with respect to the substrate's normal. 
The base pressure in the sputter tool chamber was $4.0\pm{1.0}\cdot10^{-7}$ Torr. 
The sources of the material deposited consisted of circular targets of 76.2 mm diameter  and 6.35 mm thickness (from MaTeck). The purity of the targets was 99.99\% for Ag, 99.999\% for Ge and Si, respectively. \\
Prior to sample fabrication, the sputter rate of  each individual target/element has been determined using AFM and ellipsometry measurements. The deposition rates of Ag, Ge and Si were respectively: $14.5$~nm/min; $14$~nm/min and $10$~nm/min.\\
Following the deposition, the sputtered substrates were cut into several squared shape pieces. A piece from each sample batch was annealed in a ultra high vacuum (at a pressure in the order of $10^{-7}$ Torr) oven (RTA system by Createc) in the range $100 - 550^\circ$C.

\subsubsection*{Characterisation}

\subsubsection*{Ellipsometry}
\label{sec:ellips}
A spectroscopic ellipsometer (M200-F J.A. Woollam Co. Inc., Lincoln, NE) was used to measure the optical constants of the films deposited on the silica substrates. All the measurements were performed in a wavelength range between 300 and 900 nm. All the ellipsometry measurements were performed at three different angle of incidence, from 60° to 76° with a 2° angular step.\\
As far as the measurements of the different compositions of interest are concerned, the model used consisted of a 1 mm silica substrate what was chosen in the software database for this layer and of the deposited film on top of it, modeled as a $50$~nm Bruggeman effective medium approximation (EMA) layer with the two phases (either Ag and Ge or Ag and Si) of volume fraction derived from the atomic composition of the mixture produced and being measured. The optical constants of the Ag phase used in the EMA model were taken from a measurement of the 100\% \ Ag calibration sample of known thickness. Leaving the optical constants of the semiconductor phase in the Bruggeman EMA layer as free fitting parameters, allowed the model data to fit well the experimental data. This approach is justifiable based on the fact that the crystalline fraction of this phase is not known and hence its optical constants are not well defined a priori.

\subsubsection*{Composition verification with energy-dispersive X-ray spectroscopy}
\label{sec:EDX}
The composition of the composites produced was verified by energy-dispersive X-ray spectroscopy (EDS). The device used was an EDAX Apollo X mounted on a Thermofischer scientific (former FEI) Quanta 200F operated at 20kV acceleration potential..

\subsubsection*{Transmission electron microscopy}
\label{sec:TEM}

The in-situ and conventional STEM and TEM analyses were carried out using a Thermofischer Scientifc (former FEI) Talos F200X operated at 200 kV in both, TEM and STEM operation modes. In-situ heating studies were performed with a double-tilt DensSolutions heating holder using a DEMS-D6-F-1300-S DensSolution heating chip with $<20$~nm thick SiNx membranes as a supporting foil. The material in study was sputter-deposited onto a chip and  studied within a membrane areas.\\
The FFT analyses were performed by using the TIA-software (Thermofischer Scientific) and compared to the simulated electron diffraction patterns (JEMS software package).

\section*{Acknowledgements }
This work was supported by ETH Research Grant ETH-47 18-1. The authors thank the Scientific Center for Optical and Electron Microscopy (ScopeM) and the FIRST cleanroom team at the ETH Zurich for their support. We thank N.~Spencer and the LSST for access to the ellipsometer. 

\section*{Author contributions}

H.G. and A.W. designed the research. M.S. and V.S. fabricated the samples used in the article. M.S. and H.G. performed the optical measurements. H.G., A.W. and M.S. analyzed the optical measurements. A.W. performed the Raman measurements. V.S. and H.M. performed the XRD measurements. A.S. performed and analyzed the in-situ TEM/STEM measurements. H.G. wrote the manuscript. H.G. designed the theoretical research, developed and performed the FEM simulations. R.S. suggested experiments and contributed to the interpretation. All authors contributed equally to the preparation of the manuscript.

\section*{Competing interests} The authors declare no competing interests.

\section*{Correspondence} Correspondence and requests for materials
should be addressed to Henning Galinski~(email: henning.galinski@mat.ethz.ch) and Ralph Spolenak (email: ralph.spolenak@mat.ethz.ch)

\bibliography{nature.bib}

\end{document}